\documentclass[photonics,article,accept,moreauthors]{Definitions/mdpi}

\firstpage{1} 
\pubvolume{1}
\issuenum{1}
\articlenumber{0}
\pubyear{2026}
\copyrightyear{2026}
\datereceived{ } 
\daterevised{ } 
\dateaccepted{ } 
\datepublished{ } 

\usepackage{eurosym}
\usepackage{graphicx}
\usepackage{amsmath}
\usepackage{amssymb}
\usepackage{color}

\def\cor{}

\Title{Motion of dissipative solitons in quasi-one-dimensional potentials}
\Author{Valentin Besse$^1$, Herv{\'e} Leblond$^{1,}$*, Dumitru Mihalache$^2$
and Boris A. Malomed$^{3,4}$}

\AuthorNames{Firstname Lastname, Firstname Lastname and Firstname Lastname}

\address{$^1$ \quad LUNAM Universit{\'e}, Universit{\'e} d'Angers, Laboratoire de Photonique
d'Angers, EA 4464, 2 Boulevard Lavoisier, 49000 Angers, France; herve.leblond@univ-angers.fr (H.L.)\\
$^2$ \quad Horia Hulubei National Institute of Physics and Nuclear Engineering,
RO-077125 Magurele, Bucharest, Romania; dumitru.mihalache@nipne.ro\\
$^3$ \quad Department of Physical Electronics, School of Electrical Engineering,
Faculty of Engineering, Tel Aviv University, Tel Aviv 69978, Israel; 
malomed@tauex.tau.ac.il\\
$^4$ \quad Instituto de Alta Investigaci\'{o}n, Universidad de Tarapac\'{a}, Casilla
7D, Arica, Chile}

\corres{Correspondence: herve.leblond@univ-angers.fr (H.L.)}

\abstract{We consider the motion of two-dimensional (2D) dissipative solitons and
soliton pairs in the framework of the cubic-quintic complex Ginzburg-Landau
equation with quasi-1D potentials, spatially periodic or radial-lattice
ones. The equation is solved numerically for inputs representing one or two
fundamental solitons, which are set in motion by a kick. For the quasi-1D
potential, the kick parallel to the potential axis naturally initiates free
motion of the 2D solitons, as the model remains Galilean-invariant in this
direction. Motion in the transverse direction, which may entail pulse
multiplication, commences if the kick is strong enough. Oblique kicks lead to
a straightforward superposition of the longitudinal and transverse motions.
Also considered is the collision between 2D solitons moving in the
longitudinal and diagonal directions. In the radial potential, the soliton
kicked towards the center may bounce back from the central potential hill or
pass it, producing complex trajectories and splitting of the moving soliton.}


\begin{document}
Published in  \textit{Photonics} (MDPI), \textbf{13}, 1097 (2026)
\section{Introduction}

The complex Ginzburg-Landau (CGL) equation has been the subject of extensive
research in the course of several decades, see Ref. \cite{Aranson02} for a
review of earlier works. A well-known species of modes generated by
one-dimensional (1D) CGL equations are self-trapped localized states in the
form of dissipative solitons \cite%
{Malomed87,Fauve&Thual88,Akhmediev96,SotoCrespo97,Akhmediev97,Descalzi02}. They are
supported by coexisting balances between dispersion or diffraction and
nonlinearity, as in the case of conservative solitons, and between gain and
losses.

Multidimensional solitons, including optical spatiotemporal ones in the form
of \textquotedblleft light bullets" \cite{Silberberg90,Stegeman00}, have
also drawn a great deal of interest. A crucially important issue for
multidimensional solitons is their instability against the critical or
supercritical collapse \cite{Fibich}. In particular, the stabilization of 2D
and 3D solitons, including those with embedded vorticity, can be readily
provided by competing nonlinearities, which is often represented by the
combination of cubic self-focusing and quintic defocusing terms \cite%
{Buryak95,Skarka97,Michinel1,Michinel2,Michinel3,9authors,FalcaFilho13}. The
cubic-quintic nonlinearity can also secure the stability of multidimensional
dissipative solitons in the framework of the CGL equations \cite%
{Sakaguchi,Grelu05,SotoCrespo06,Mihalache06}, including vortex
(\textquotedblleft spinning") doughnut-shaped ones, see also reviews \cite%
{Mihalache08,Malomed22}.

Another generic possibility to create stable 2D and 3D solitons is offered
by models including spatially periodic (lattice) potentials. most frequently
a periodic one. The potentials can represent laser cavities created in
photonic crystals, induced by a spatially periodic modulation of the
refractive index in the transverse directions \cite{PhotCryst1,PhotCryst2},
as well as a Bose-Einstein condensate (BEC) of excitons trapped in an
optical lattice~\cite{Dalfovo99,excitons}. Solutions of the Gross-Pitaevskii equations
with spatially and radially periodic potentials have been obtained in the
form of stable vortex dissipative solitons~\cite{Mihalache10a}. In the
spatiotemporal (effectively 3D) domain, relevant models include quasi-2D
lattice potentials. Stable fundamental (zero-vorticity) and quasi-stable
vortex solitons for such models were found in Refs. \cite%
{Baizakov04,Mihalache04} and \cite{Leblond07}, respectively (see also the review
\cite{Mihalache10b}). A dissipative axisymmetric lattice, in the form of a
radially periodic modulation of the gain and loss, rather than the
reflective index, was considered too \cite{Skarka10,Skarka14}. In the latter
case, the respective CGL equation produces localized star-shaped patterns
with nonzero angular momentum but zero topological charge. Other
low-dimensional potentials have been considered too, with razor, dagger and
needle shapes~\cite{Liu10}, as well as azimuthally varying and conical~ones
\cite{Liu11}.

Motion of solitons in the CGL model with a 2D transverse potential was
considered in Ref. \cite{Besse_2013}. It was found that the fundamental
soliton, hopping between potential cells, leaves arrayed patterns of
different sizes in its wake. The motion of dipole and vortex dissipative
solitons in the same model demonstrates similar pattern-formation scenario
\cite{Besse_2014,Mihalache15}. It is worthy to note that the Newton's-cradle
dynamics was observed in the train of secondary solitons, generated by the
moving primary one. A similar dynamical regime was predicted in soliton
trains hosted by optical fibers~\cite{Driben13}.

Interactions between dissipative solitons determine the two- and
multi-soliton dynamics. In the framework of the CGL models, bound states of
dissipative solitons were predicted long ago~\cite{Afanasjev97,Akhmediev97b},
 and observed in fiber lasers \cite{Gong01,Hideur03}, see also the recent
review \cite{Cui26}. Other types of the interaction, such as their merging
and mutual destruction, were predicted too~\cite{Besse_2013,Besse_2014} and
observed in a passive optical cavity~\cite{Jang16}.

In the present work, we consider the motion of dissipative solitons in the
2D cubic-quintic CGL model with quasi-1D potentials. In Sections II and III, we
address, respectively, the dynamics of a single fundamental soliton and
soliton pair in a quasi-1D potential. The subject of Section IV is the
dynamics of a dissipative soliton in the potential in the form of a radial
lattice. The paper is concluded by Section V.

\section{Motion of the fundamental soliton in the quasi-1D potential}

We consider the cubic-quintic CGL equation, for amplitude $u\left(
x,y;t\right) $ of the optical field, in its usual form,
\begin{equation}
\frac{\partial u}{\partial t}=-\left[ \delta +iV(x,y)\right]
u+(i+\varepsilon )|u|^{2}u-(i\nu +\mu )|u|^{4}u+\frac{i}{2}\Delta u,
\label{cgl}
\end{equation}%
where $t$ is the propagation distance in optics, $\Delta $ is the Laplacian
operator, and the quasi-1D lattice potential is
\begin{equation}
V(x,y)=-V_{0}\cos \left( 2y\right) ,  \label{V}
\end{equation}%
whose period is fixed to be $\pi $ by rescaling.
\cor{This period is commensurate with the numerical box size.}
\cor{The model applies, as a general one, to two-dimensional laser
cavities~\cite{Skarka97,Sakaguchi,Grelu05,SotoCrespo06,Malomed22,excitons,Mihalache10a,Skarka14,Liu10,Liu11,Besse_2013,Besse_2014,Mihalache15}. 
In these numerous previously published works, physical conditions for the applicability of 
such models have been considered in detail, therefore we refer the reader to those works for further information.}
\cor{Diffusion can be added to the model by replacing the imaginary coefficient in front of the Laplacian in Eq. (\ref{cgl}) with a complex one, see the report~\cite{Rosanov26}.
 However, diffusion of light does not occur in laser cavities, and hence will not be
 accounted for here.}
The scaling is also used to fix the cubic self-focusing and diffraction
coefficients to be $1$. Generic results are presented below by varying the depth
$V_{0}$ of the lattice potential (\ref{V}), for fixed characteristic values
\begin{equation}
\delta =0.4,\varepsilon =1.85,\nu =0.1,~\mathrm{and~}\mu =1,  \label{coeff}
\end{equation}%
of the linear loss, cubic gain, quintic defocusing, and quintic loss
coefficients, respectively.

\begin{figure}[H]
\includegraphics[width=8cm]{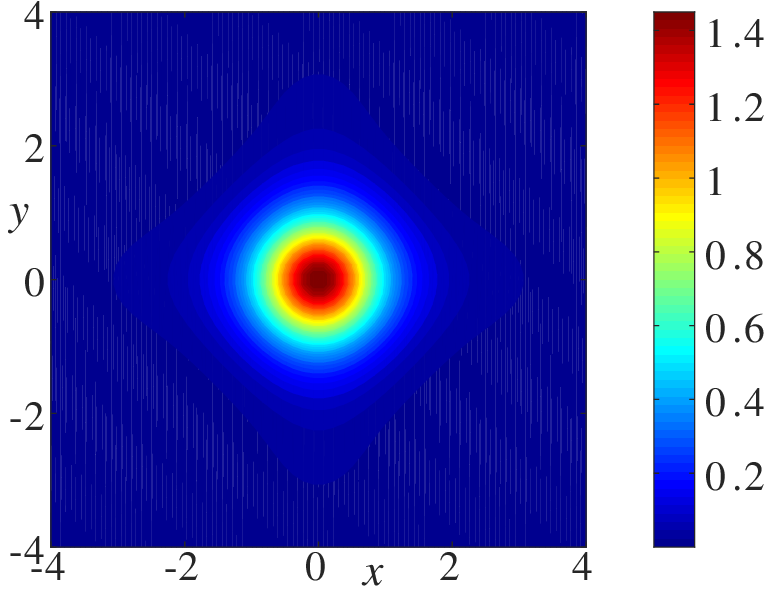}
\caption{The stable fundamental dissipative 2D soliton produced by the
numerical solution of Eq. (\protect\ref{cgl}) with $V_{0}=1$ in the quasi-1D
potential (\protect\ref{V}) and coefficients (\protect\ref{coeff}). Plotted is $|u|$,
representing the solution as
$ u(x,y,t)=\left\vert u\left(  x,y\right) \right\vert \exp \left( -i\omega t\right) $
with $\omega=-0.377$.
\label{fig_fond}}
\end{figure}

Simulations of Eq. (\ref{cgl}) were performed by means of the standard
fourth-order Runge-Kutta algorithm with an adaptive step for variable $t$,
while the second derivatives with respect to $x$ and $y$ were computed using
five-point finite-differences formulas, with periodic boundary conditions.
\cor{We use a $292\times 292$ numerical grid centered at (0,0) with 
a box size given by $\max(x)=\max(y)=21.991=7\pi$, for the periodic potential.
For the circular potential, the grid is $280\times 280$, and the box size
$\max(x)=\max(y)=20.955$, the center being still (0,0). In this case,
a few computations are performed with a smaller grid of
$140\times 140$, with $\max(x)=\max(y)=10.477$.}
An example of the stable fundamental soliton produced by the numerical
scheme for $V_{0}=1$ is
displayed in Fig. \ref{fig_fond}. To set the dissipative soliton in motion,
we apply the kick to it:
\begin{equation}
u\left( x,y\right) \rightarrow u\left( x,y\right) \exp \left( i\mathbf{k}%
_{0}\cdot \mathbf{r}\right) ,\ \mathbf{k}_{0}=k_{0}(\cos \theta ,\sin \theta
),~\mathbf{r}=\left( x,y\right) ,  \label{kick}
\end{equation}%
and use expression (\ref{kick}) as the input for simulations of Eq. (\ref%
{cgl}). Systematic results are produced by varying the kick's strength $%
k_{0} $ and direction $\theta $.

\begin{figure}[H]
\includegraphics[width=8cm]{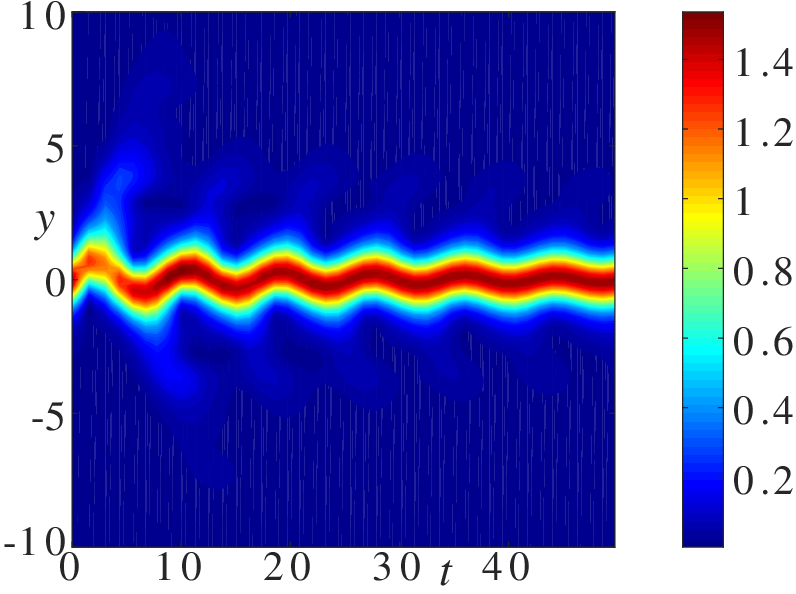}
\caption{Transient oscillations of the soliton in the deep potential (%
\protect\ref{V}), with $V_{0}=1$, \ initiated by a relatively strong
transverse kick (\protect\ref{kick}) with $k_{0}=1.3$ and $\protect\theta =%
\protect\pi /2$. eventually stops. Plotted is $|u(t,y)|$ in cross section $%
x=0$.
\label{fig_osc}}
\end{figure}

Figure \ref{fig_osc} demonstrates that a sufficiently deep potential (\ref{V}%
) (in particular, with $V_{0}=1$) keeps the dissipative soliton, kicked
across the potential ($\theta =\pi /2$ in Eq. (\ref{kick})). In this case,
the kick initiates gradually decaying oscillations of the soliton in the
verticall direction. This outcome occurs for $\theta =\pi /2$, $V_{0}=1$,
and $0.1\leqslant k_{0}\leqslant 1.81$.
The upper bound of this outcome is $k_0 = 1.53$
 for the potential with an intermediate strength, $V_0 = 2/3$, and $ k_0 = 0.6$ for $V_0 = 0.2$.
Naturally, the period $T$ of the damped
oscillations increase for a weaker potential: for instance, the kick with $%
k_{0}=0.5$ gives rise to the oscillations with $T=8$, $12$, and $25$ for $%
V_{0}=1$, $0.66$, and $0.2$, respectively.
\cor{These oscillations closely resemble the trapped waves reported for 
solitons interacting with an external force field in KdV-type models 
with localized~\cite{Flamarion_Chaos} 
and periodic~\cite{Flamarion_Math_kdv} potentials, except that there
the oscillation amplitude grows in time while here it is damped. 
}

\begin{figure}[H]
\includegraphics[width=8cm]{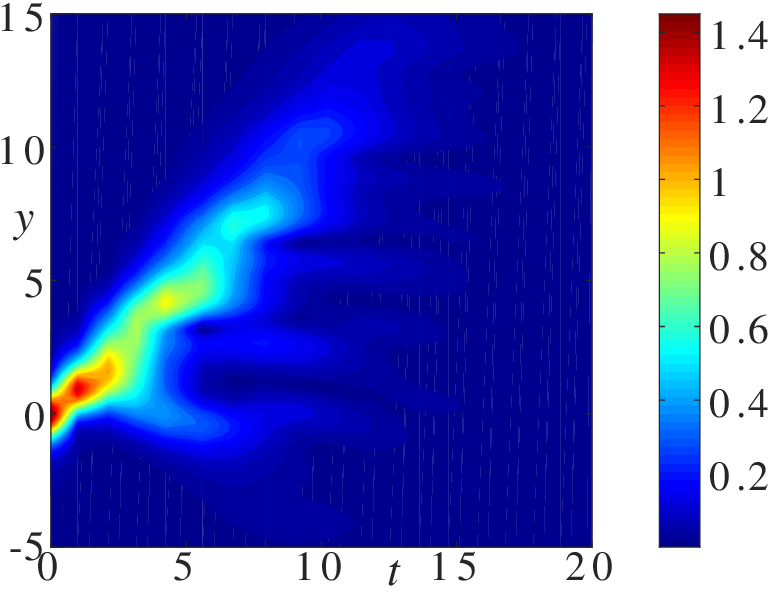}
\caption{The creation of the additional soliton in the adjacent potential
trough by the application of a strong kick (\protect\ref{kick}), with $%
k_{0}=2.2$ and $\protect\theta =\protect\pi /2$, to the potential (\protect
\ref{V}) with $V_{0}=1$, followed by quick decay of the entire pattern.
Plotted is $|u(t,y)|$ in cross section $x=0$.
\label{fig_vanish}}
\end{figure}

For the same relatively deep quasi-1D potential as in Fig. \ref{fig_osc},
with $V_{0}=1$, a stronger transverse kick (\ref{kick}), with $\theta =\pi
/2 $ and $1.82\leqslant k_{0}\leqslant 2.407$, applied to the initial
soliton overcomes the potential barrier, by creating  an additional
soliton-like spot in the adjacent trough, in the direction of the kick.
However, as shown in Fig. \ref{fig_vanish}, the so produced double soliton
is unstable, disappearing after passing a small propagation distance $t$.
For a more shallow quasi-1D potential, with $V_{0}=0.66$ in Eq. (\ref{V}),
the application of the kick produces essentially the same outcome in the 
interval $1.54\leq k_{0}$ $\leq 1.81$.

\begin{figure}[H]
\includegraphics[width=8cm]{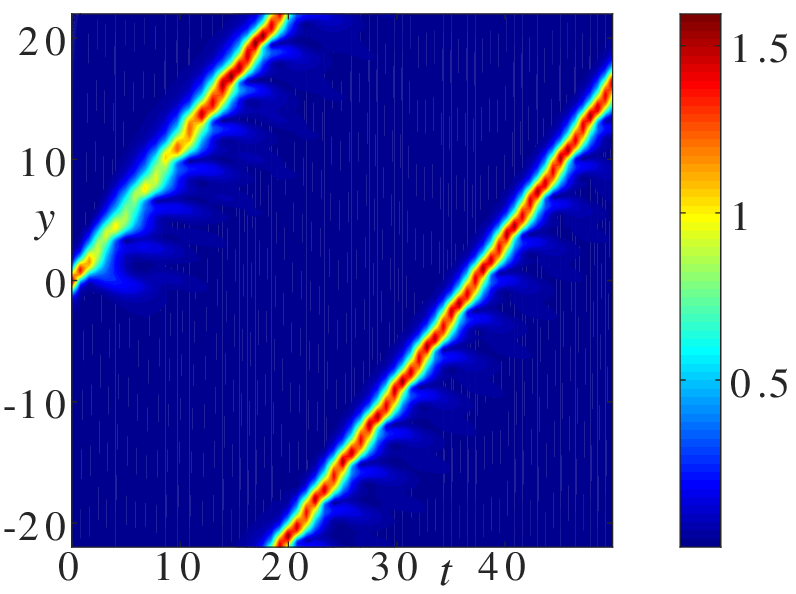}
\caption{The strong transverse kick (\protect\ref{kick}), with $k_{0}=2.5$
and $\protect\theta =\protect\pi /2$. applied to the relatively deep
quasi-1D potential (\protect\ref{V}), with $V_{0}=1$, sets the dissipative
soliton in the state of quasi-steady motion. Plotted is $|u(t,y)|$ in cross
section $x=0$.
\label{fig_propag}}
\end{figure}

Finally, Fig. \ref{fig_propag} demonstrates that the strongest kicks
considered in this work, in the interval of $2.408\leq k_{0}\leq 3$, if
again applied to the dissipative soliton under the the action of the
quasi-1D potential (\ref{V}) leads to another output: after a transient
stage of attenuation and revival, the initial soliton is converted into a
quasi-steadily-moving one. {Naturally, its velocity increases with the kick's
magnitude $k_{0}$, while the duration of the transient stage reduces.} For a
weaker potential, the steady-propagation regime commences from smaller
values of $k_{0}$ -- e.g., it occurs at $k_{0}\geq 1.82$ for $V_{0}=0.66$.

\begin{figure}[H]
\includegraphics[width=8cm]{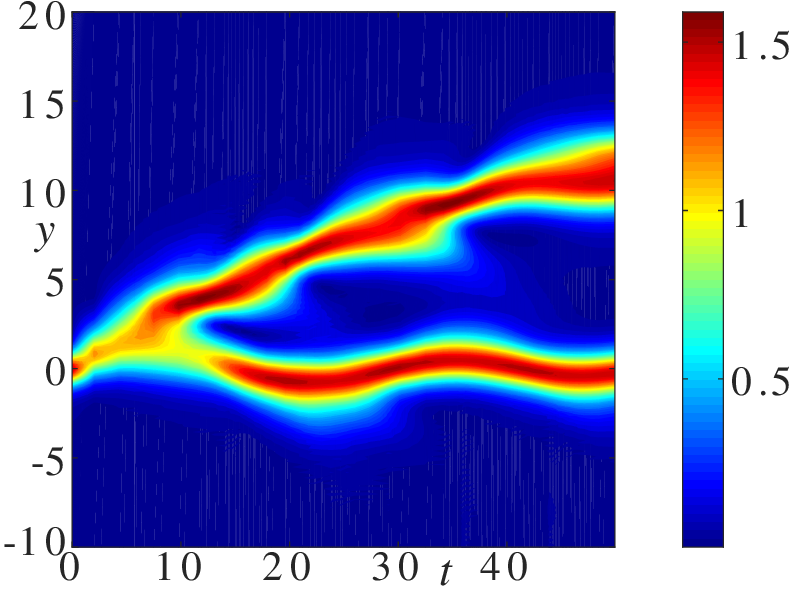}
\caption{The duplication of stable dissipative soliton by kick (\protect\ref%
{kick}) with $k_{0}=0.94$ and $\protect\theta =\protect\pi /2$. applied to
the shallow potential with $V_{0}=0.2$ in Eq. (\protect\ref{V}). Plotted is $%
|u(t,y)|$ in cross section $x=0$.
\label{fig_dupli}}
\end{figure}

The 2D\ CGL model based on Eq. (\ref{cgl}) with a shallow quasi-1D potential
(\ref{V}) (e.g., with $V_{0}=0.2$) produces different phenomenology for
kicked dissipative solitons. The destruction regime, illustrated in Fig. \ref%
{fig_vanish}, does not occur in this case. Instead, the kick sets the
soliton in a state of low motion, without causing its decay. For instance,
the transverse ($\theta =\pi /2$) kick with $k_{0}=0.70$ and $0.78$ imparts
small velocities to the soliton, with which it passes, respectively, a
single cell or three cells of the lattice within the full simulation, up to $%
t=50$. In the same case of $V_{0}=0.2$, stronger kicks, in the interval of
$0.79\leqslant k_{0}\leqslant 0.94$
lead to duplication of the soliton, somewhat similar to the initial stage
observed in Fig. \ref{fig_vanish} (for $V_{0}=1$), but it is not followed by
disappearance of the solitons.
For $k_0\leqslant 0.84$ two solitons persist.
 Counterintuitively, the observed motion is faster for $%
k_{0}=0.79$ than for larger magnitudes of the kick.
From $k_0=0.85$  up to $k_{0}=0.89$, the
rear spot is eventually reabsorbed by the front one, while at $k_{0}>0.89$
both solitons persist, the front one moving steadily and the rear one
demonstrating slow oscillations, see Fig. \ref{fig_dupli}. 
\cor{Note that soliton splitting has also been reported for the average field
of Benjamin-Ono solitons affected by random perturbations \cite{FlamarionMath}.
}
Eventually, at $%
k_{0}\geqslant 0.95$ the rear soliton vanishes so that the simulations
produce a single moving soliton.

The above results are produced for the transverse kick, with $\theta =\pi /2$
in Eq. (\ref{kick}). The longitudinal kick, with $\theta =0$, immediately
sets the soliton in free motion, as Eq. (\ref{cgl}) keeps Galilean
invariance in the $x$ direction, cf. Ref. \cite{Sakaguchi}. An oblique kick
(e.g., with $\theta =\pi /4$) leads to straightforward superposition of the
motions in the $y$ and $x$ directions.

\begin{figure}[H]
\includegraphics[width=8cm]{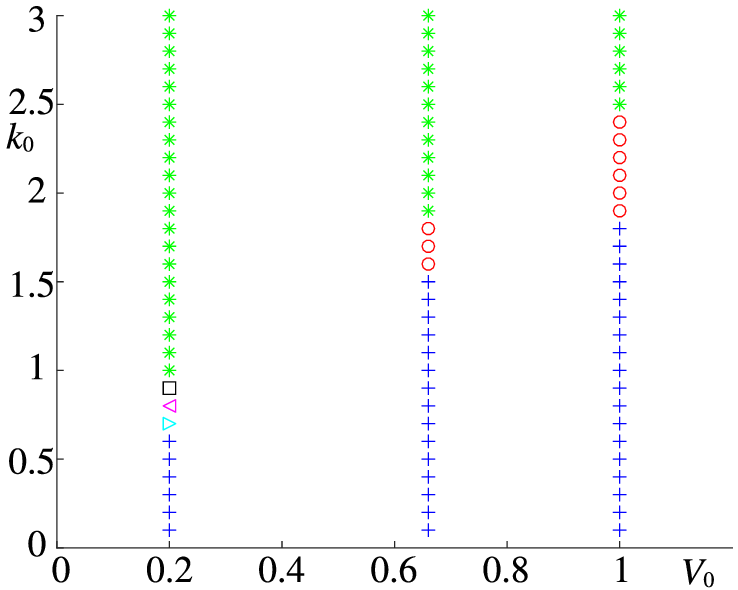}
\caption{The different outcomes of kicking the single 2D soliton in
the transverse direction ($ \theta =\pi /2$ ), versus $ V_0$ and $ k_ 0$.
$\textcolor{blue}+$: damped oscillations; $\textcolor{red}\circ$:
disappearance of the soliton; $\textcolor{green}\ast$:
soliton motion; $\textcolor{cyan}\triangleright$: slow motion;
$\textcolor{magenta}\triangleleft$: two-soliton final state;
$\textcolor{black}\square$: duplication and slow motion of the front soliton.
\label{chart}}
\end{figure}

Figure \ref{chart} shows the different outcomes of kicking the single 2D soliton in
the transverse direction ($ \theta =\pi /2$ ), in the plane
of two parameters $ V_0$ and $ k_ 0$,
while other parameters are fixed as per Eq. (\ref{coeff}).

\section{The quasi-1D potential: dynamics of soliton pairs}

To study the two-soliton dynamics, we prepare an input containing two solitons
in adjacent troughs of the quasi-1D potential (\ref{V}), with identical
phases. Then the soliton pair is set in motion by the application of kick (%
\ref{kick}) to both solitons. The longitudinal kick with $\theta =0$ (at
least, in the interval of $0.1\leq k_{0}\leq 5$) creates a pair of
synchronously moving solitons, irrespective of the fact if the solitons form
a true bound state or not.

More interesting are effects produced by the transverse kick, with $\theta
=\pi /2$. If the kick is weak, $0.1\leqslant k_{0}\leqslant 1.4$,
the widths and vertical positions of the two solitons oscillate, but the
solitons do not cross the potential barrier.

\begin{figure}[H]
\includegraphics[width=8cm]{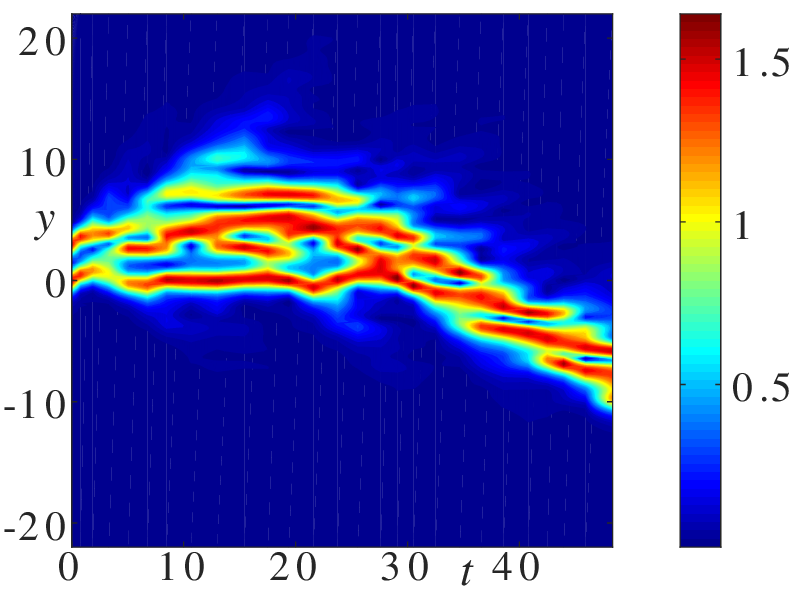}
\caption{The kicked soliton pair, with $V_{0}=1$, $k_{0}=1.5$, and $\protect%
\theta =\protect\pi /2$, creates another pair, moving forward, and then
inverts the direction of motion, while the additional pair disappears.
Plotted is $|u(t,y)|$ in cross section $x=0$.
\label{fig_back}}
\end{figure}

In the interval of $1.5\leqslant k_{0}\leqslant 1.8$, the stronger kick
initiates a complex dynamics, including soliton multiplication. At $k_{0}=1.5$%
, the kicked solitons first move forward, leaving copies of themselves in
place, thus yielding a train of four spots. Then, they move backward, in the
form of a pair of quasi-soliton, as shown in Fig. \ref{fig_back}. While the
backward propagation also occurs at $k_{0}=1.51$, in the interval of $%
1.52\leq k_{0}\leq 1.55$ the initial soliton pair propagates forward,
generating a quiescent soliton train in its wake, as seen in Fig. \ref%
{fig_train}.
\begin{figure}[H]
\includegraphics[width=8cm]{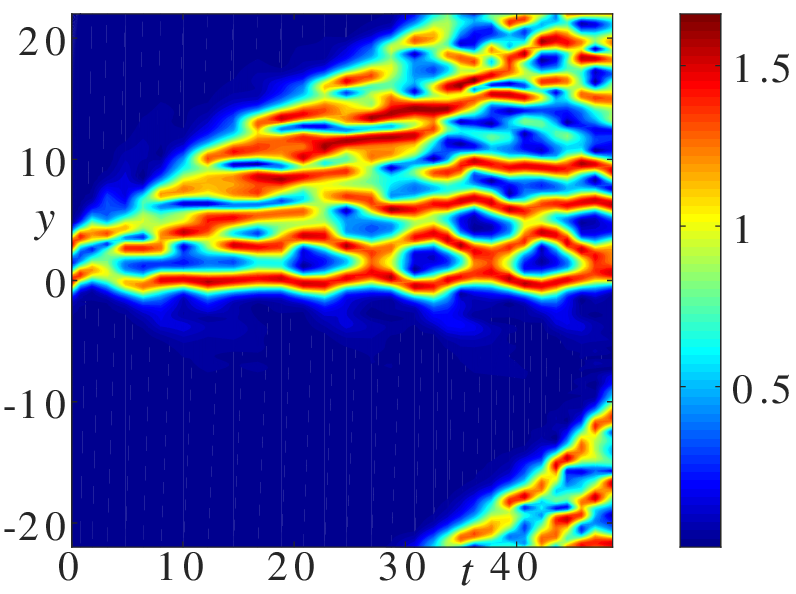}
\caption{The forward motion of the soliton pair generates a train of
quiescent solitons, as shown here for $k_{0}=1.54$, $\protect\theta =\protect%
\pi /2$, and $V_{0}=1$. Plotted is $|u(t,y)|$ in cross section $x=0$.
\label{fig_train}}
\end{figure}
\ \ \

At $k_{0}\geq 1.56$, empty sites appear between the moving soliton and the
train. The train length increases with $k_{0}$, reaching a maximum value
corresponding to $10$ solitons in the interval of $1.71\leq k_{0}\leq $ $%
1.73 $. Then, the length decreases, while the number of moving solitons is
reduced to one. Eventually, the train is reduced to a single soliton at $%
1.9\leq k_{0}\leq 2.5$. At still larger values of $k_{0}$, the fixed soliton
disappears and we are left with a single moving soliton. This behavior has
been checked up to $k_{0}=3.5$.

\begin{figure}[H]
\includegraphics[width=8cm]{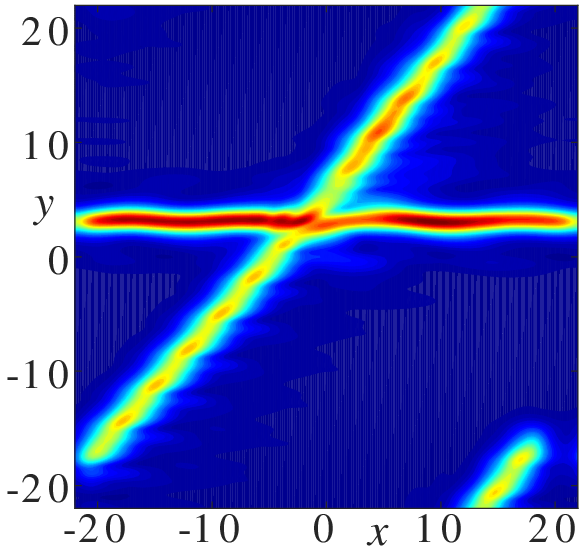}
\caption{An example of the quasi-elastic collision between solitons initially
moving along\ the diagonal, $y=x$, and one moving along the $x$-axis.
 Plotted is of $%
\protect\int_{27.2}^{43.6}|u(x,y;t)|dt$, allowing the visualization of the
trajectories in the $\left( x,y\right) $ plane. The kick (\protect\ref{kick}%
) with $k_{0}=3.5$ and $\protect\theta =\protect\pi /4$ was initially
applied to the moving solitons. The depth of the underlying quasi-1D
potential is $V_{0}=1$. The trajectory of the moving soliton reappears due
to periodic boundary conditions.
\label{fig_coll}}
\end{figure}

As well as in the case of the single-soliton dynamics, the result of the
application of the oblique kick (\ref{kick}), with $\theta \neq \pi /2$, can
be understood as  the superposition of the dynamics in the transverse ($y$%
) direction and free motion along $x$.
A new possibility, displayed in Fig. \ref{fig_coll} is the collision between solitons
moving in different directions, as a result of the application of
an oblique  ($\theta =\pi /4$)
relatively strong kick (\ref{kick}), $2.6\leqslant k_{0}\leqslant 3.5$:
one soliton moves along the $x$-axis, and the other one in the diagonal direction.
The background potential (\ref{V}) is relatively strong too, with $V_{0}=1
$. It is seen that the collision is quasi-elastic, although a conspicuous
excitation is generated in the soliton moving in the diagonal direction.

%

\section{The circular potential}

In this Section we consider the 2D CGL equation (\ref{cgl}) with the
potential representing a radial lattice,
\begin{equation}
V=V_{0}\cos ^{2}\left( \frac{r}{2}\right) ,~r=\sqrt{x^{2}+y^{2}}\text{.}~
\label{radial}
\end{equation}%
We use the same parameter values (\ref{coeff})\ as fixed above, and set $%
V_{0}=1$. The potential has minima at circumferences $r=\pi \left(
1+2n\right) $. \cor{The periodic boundary conditions strongly perturb
 the propagation when the soliton goes too far from the center of the potential.
  The results of the computations are disregarded as soon as this occurs.}

\begin{figure}[H]
\includegraphics[width=8cm]{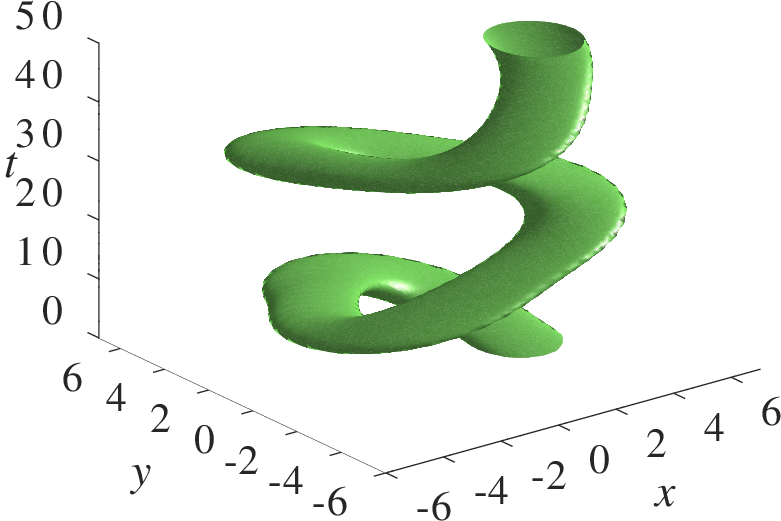}
\caption{The circular motion of the fundamental dissipative solitons
initiated by the kick (\protect\ref{kick}), wirh $k_{0}=1.3$ and $\protect%
\theta =\protect\pi /2$. Plotted is the equidensity surface, $\left\vert
u\left( x,y;t\right) \right\vert =0.5$, which is produced by the numerical
solution of Eq. (\protect\ref{cgl}) with coefficients (\protect\ref{coeff})
and potential (\protect\ref{radial}) with $V_{0}=1$.
\label{fig_rotat}}
\end{figure}

First we consider a fundamental (zero-vorticity) dissipative soliton trapped
in the first circular potential trough, with the minimum at $r=\pi .$
The soliton is set in motion along the trough by
applying kick (\ref{kick}) to it in the tangential direction, with $\theta
=\pi /2$. In the interval of the kick's strength, $0.1\leqslant
k_{0}\leqslant 1.5$, the soliton performs steady circular motion (resembling
the \textquotedblleft rotary solitons" known in conservative models \cite{YK}%
), as shown in Fig. \ref{fig_rotat}.
\vspace{1cm}

\begin{figure}[H]
\includegraphics[width=8cm]{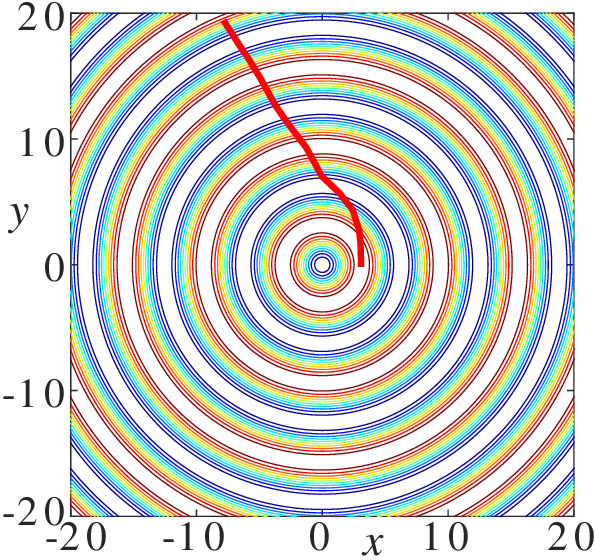}
\caption{The motion of the soliton across the radial lattice, due to the
action of the large centrifugal force, corresponding to the kick's magnitude
$k_{0}=1.6$. Plotted is 
the trajectory of the soliton's maximum in the $\left(
x,y\right) $ plane. 
\label{fig_follow}}
\end{figure}

 Its speed increases with $k_{0}$, but
eventually saturates for largest values of $k_{0}$. {At $k_{0}=1.5$, the moving
soliton emits a fleeting spot in the neighbor channel.} 
For larger $k_{0}$ (e.g., $k_{0}=1.6$), Fig. \ref{fig_follow} demonstrates
that 
the soliton, under the action the centrifugal force, moves across the radial
lattice potential. 
 From $k_{0}=1.7$ to $2.3$, 
 the trajectory is still deviated by the potential, but the
deviation reduces as $k_{0}$ increases. From $k_{0}=2.4$ to $3$, the
trajectory, driven by the strong centrifugal force, 
 remains almost
rectilinear up to the end of the simulation.
We stop the simulation as the spot touches the boundary, as shown in Fig. \ref%
{fig_gerade}.
\begin{figure}[H]
\includegraphics[width=8cm]{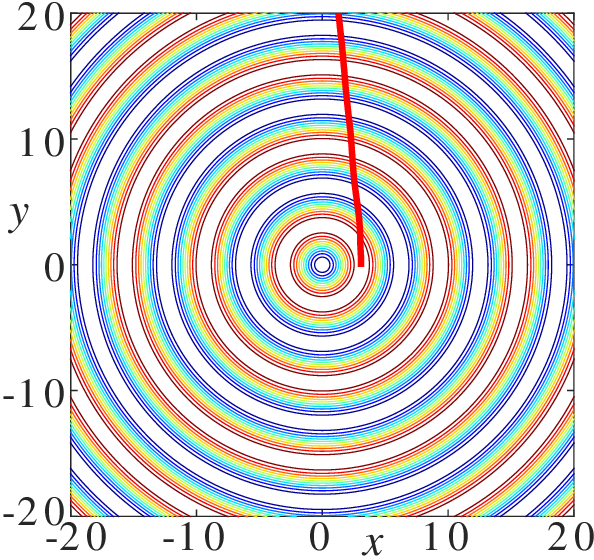}
\caption{The soliton's trajectory is nearly straight in the case of a very
strong kick, $k_{0}=2.8$. Plotted is 
the trajectory of the soliton maximum
in the $\left( x,y\right) $ plane. 
\label{fig_gerade}}
\end{figure}

\begin{figure}[H]
\includegraphics[width=8cm]{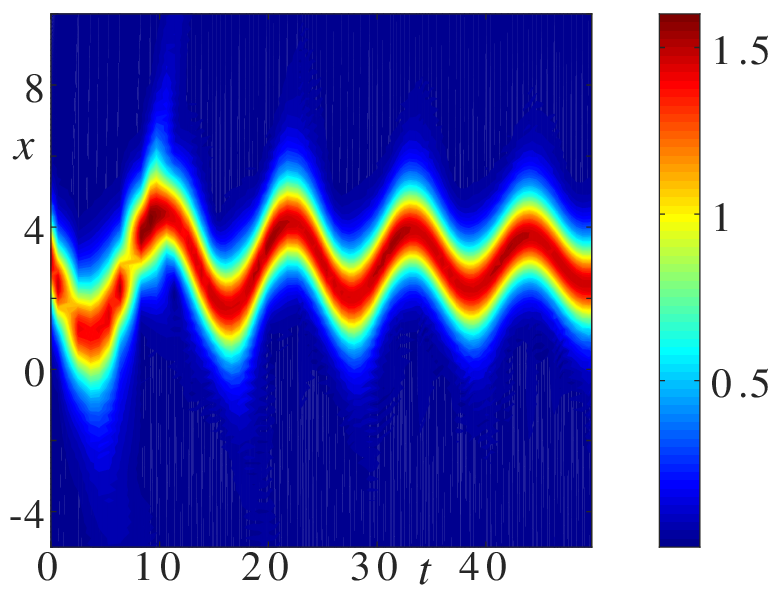}
\caption{The soliton oscillates in the transverse (radial) direction
initiated by the application of the kick with $k_{0}=1.1$ towards $r=0$
(across the circular potential). Plotted is $|u(x,t)|$ in cross section $y=0$. 
\label{fig_damped}}
\end{figure}

Next we consider the action of the kick directed towards the center of the
potential (with $\theta =\pi $). For $0.1\leqslant k_{0}\leqslant 1.1$, we
observe damped oscillations in the radial direction (transversely to the
circular potential trough), see Fig. \ref{fig_damped}. Deformations of the
solitons are visible too, especially for larger values of $k_{0}$.

\begin{figure}[H]
\begin{adjustwidth}{-\extralength}{0cm}
\centering
\subfloat[\centering]{\includegraphics[width=6cm]{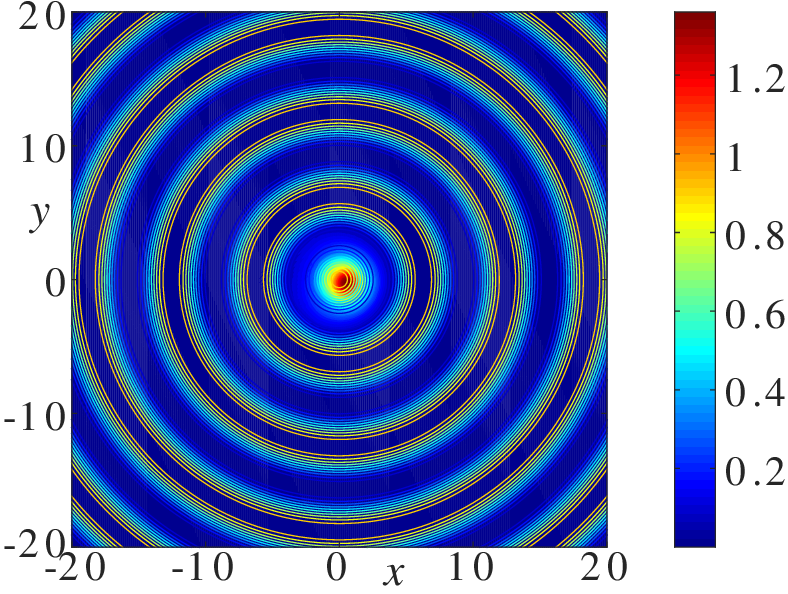}}\hspace{1cm} %
\subfloat[\centering]{\includegraphics[width=6cm]{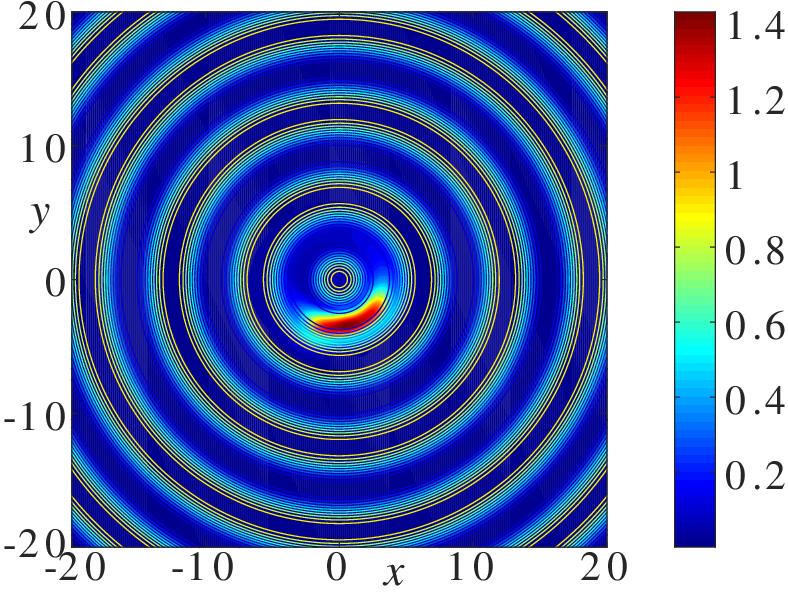}} \vspace{5mm}\\%
\subfloat[\centering]{\includegraphics[width=6cm]{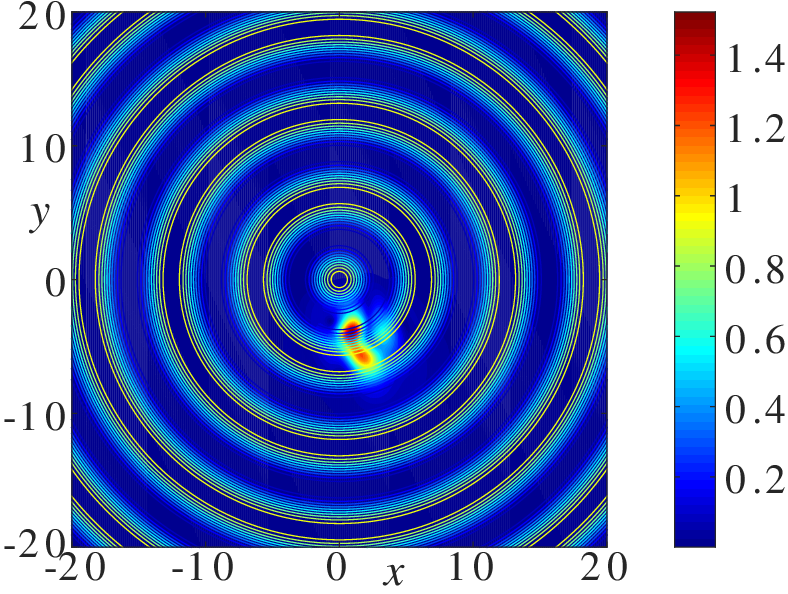}}\hspace{1cm} %
\subfloat[\centering]{\includegraphics[width=6cm]{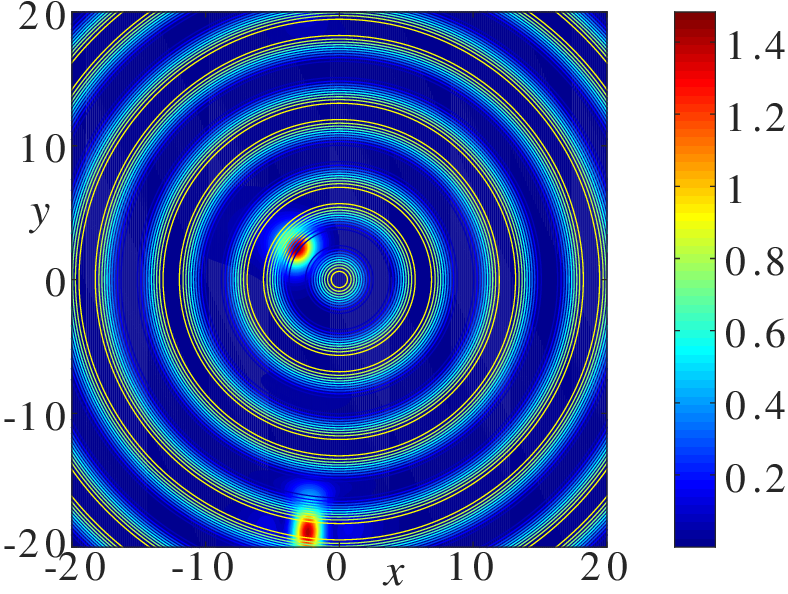}}
\end{adjustwidth}
\caption{The soliton kicked towards the center, by the push (\protect\ref%
{kick}) with $k_{0}=1.2$ and $\protect\theta =\protect\pi $, bounces off the
central potential hill [(\textbf{a}), $t=3.65$], takes a crescent shape [(\textbf{b}), $t=8.66$%
], duplicates 
[(\textbf{c}), $t=12.45$], then one soliton moves around the potential trough
while the other goes straight away 
 [(\textbf{d}), $t=23.53$]. Plotted is $|u(x,y)|$ in the
$\left( x,y\right) $ plane.
\label{fig_lune}}
\end{figure}

A stronger inward kick, with $k_{0}=1.2$, sends the dissipative soliton
towards the center. It bounces back from the central potential hill, warps
into a crescent, and then
splits into two solitons. One of them  
starts circular motion around the potential trough,
while the other one goes straight away, see Fig. \ref{fig_lune}.
The choice of a particular direction of the circular motion implies the
occurrence of spontaneous symmetry breaking. 

\begin{figure}[tbp]
\begin{adjustwidth}{-\extralength}{0cm}
\centering
\subfloat[\centering]{\includegraphics[width=6cm]{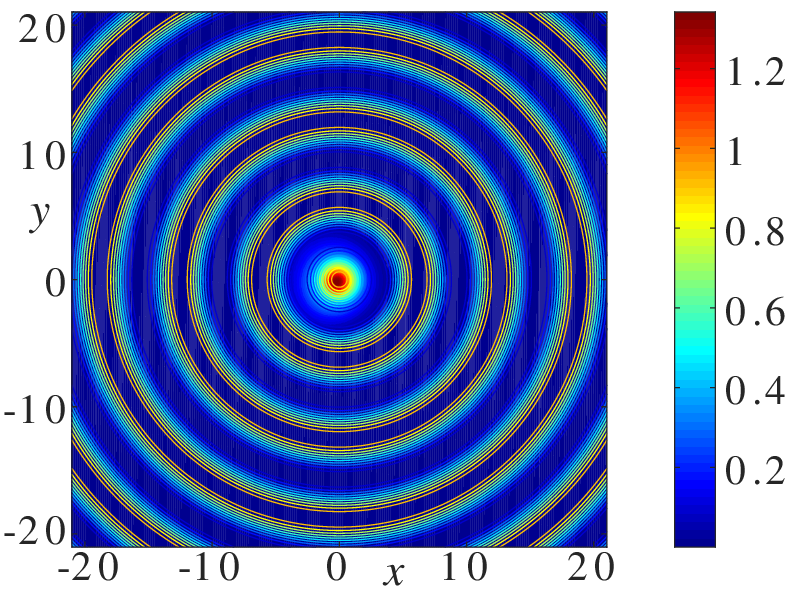}}\hspace{1cm} %
\subfloat[\centering]{\includegraphics[width=6cm]{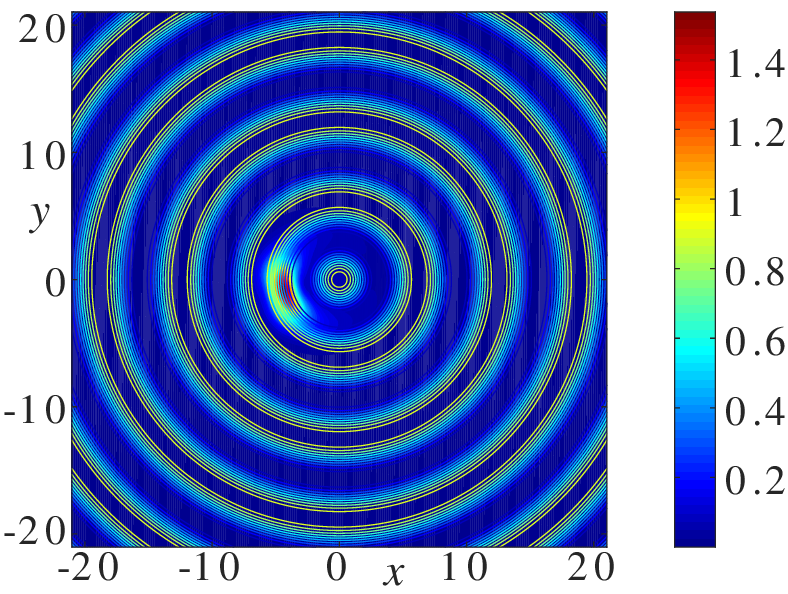}} \vspace{5mm}\\%
\subfloat[\centering]{\includegraphics[width=6cm]{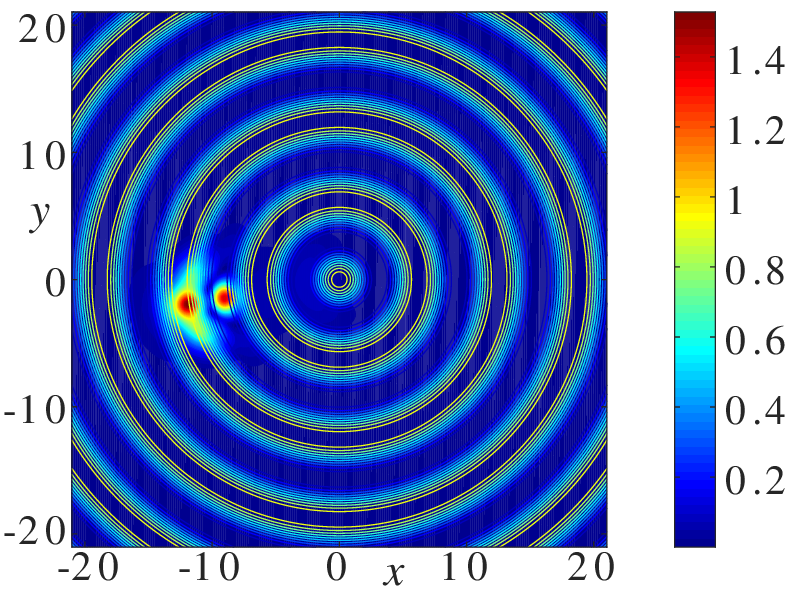}}\hspace{1cm} %
\subfloat[\centering]{\includegraphics[width=6cm]{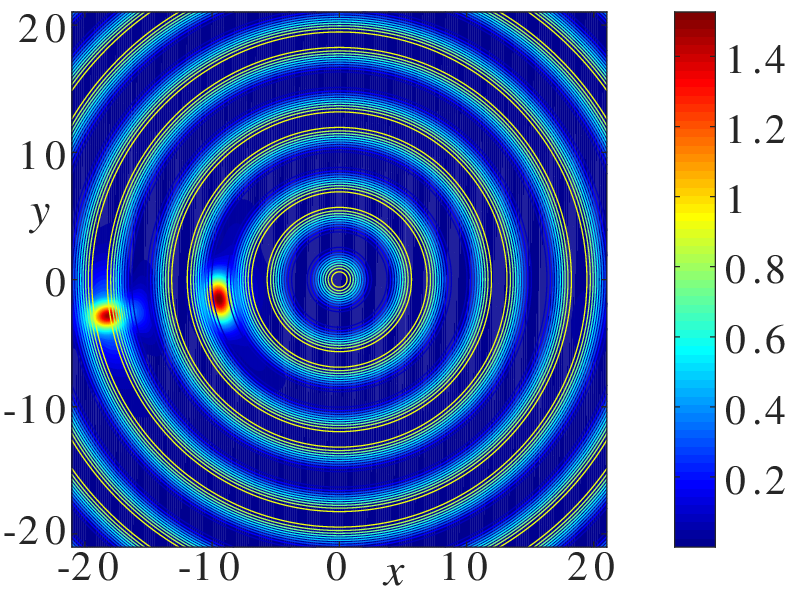}}
\end{adjustwidth}
\caption{The soliton kicked inward as per Eq. (\protect\ref{kick}), with $%
k_{0}=1.5$ and $\protect\theta =\protect\pi $, passes the central potential
hill [panel (\textbf{a}), corresponding to $t=2.88$] and stops at the first annular potential barrier, spreading into a
crescent-shaped structure [(\textbf{b}), $t=6.12$]. Then. the
crescent splits in two fragments [(\textbf{c}), $t=15.52$], one
 of which stays motionless in the annular
potential trough, while the second one goes straight [(\textbf{d}), $t=20.77$],
in a direction oblique with respect to the $x$-axis, 
 featuring
symmetry breaking. Plotted is $|u(x,y;t)|$ in the $\left(
x,y\right) $ plane.
\label{fig_cross}}
\end{figure}

The evolution initiated by the strong inward kick, with $k_{0}\geq 1.5$, is
represented by Fig. \ref{fig_cross}. In that case, the kicked soliton
acquires enough momentum to pass the central potential hill, gradually
losing the momentum in the course of motion. It continues straight until the
momentum is insufficient to pass a particular circular barrier, where it
stops, spreading into a crescent. Naturally, the distance passed before the
stoppage increases with $k_{0}$. In the case of $k_{0}=1.5$, Fig. \ref%
{fig_cross} demonstrates that, after the kicked soliton passed the central
hill, it stops at the first annular potential barrier and spreads into a
crescent, which splits into two spots. One of them stays motionless in
 the annular channel, and the other one keeps moving
straight. Its direction is not parallel to the $x$-axis,
 featuring symmetry breaking.
For a higher inward kick, the initial spot continues propagating in the left
direction up to the boundary of the numerical box.

\section{Discussion}

In the quasi-1D lattice potential, the 2D dissipative soliton set in motion
by the transverse kick (\ref{kick}), with $\theta =\pi /2$, stays in the
original potential trough if the kick's strength $k_{0}$ is small. On the
contrary, the soliton kicked with large $k_{0}$ moves freely across the
potential lattice. At intermediate values of $k_{0}$, more complicated
dynamics occurs, with jumps between adjacent troughs and soliton
duplication. Regarding the soliton pair, only a single freely moving soliton
can be produced by the kick, and the dynamics of the intermediate regime is more
complex, involving the formation of a soliton train.

When the kick is applied parallel to the quasi-1D potential ($\theta =0$ in
Eq. (\ref{kick})), both the single soliton and soliton pair move freely
along the direction that admits the Galilean invariance. An oblique kick ($%
0<\theta <\pi /2$) typically initiates motion that is a straightforward
superposition of the two transverse and longitudinal modes. A separately
considered case was the collision between two dissipative solitons moving in
the diagonal and longitudinal directions.

In the radial lattice potential, the application of a relatively weak kick
in the tangential directions initiates motion along an annular potential
trough. However, a stronger tangential kick produces a fast motion with a
large centrifugal force, which drives the solitons along complex
trajectories crossing the radial potential barrier.

When the strong kick is directed towards the center of the radial lattice,
it pushes the dissipative soliton across the central potential hill.
Eventually, the soliton stops at a certain annular potential barrier,
spreading into a crescent structure, which, in turn, may split into 
two fragments, one of which stays motionless, 
 while the second one keeps moving in a slightly oblique 
direction. 
The outcome of the
evolution features spontaneous symmetry breaking.

As an extension of the analysis, it may be interesting to explore the
evolution of 2D dissipative solitons carrying embedded vorticity. In
particular, the longitudinal motion along the free direction of the quasi-1D
potential may be perturbed in the transverse direction by an effective
\textquotedblleft Magnus force".

\authorcontributions{Conceptualization, B.M. and V.B.;
 software, V.B.; investigation, V.B. and H.L.; writing, H.L., B.M. and D.M.  All authors have read and agreed to the published version of the manuscript.}

\funding{This research received no external funding.}

\dataavailability{Data will be available on request.}

\conflictsofinterest{The authors declare no conflicts of interest.}

\reftitle{References}

\end{document}